\documentstyle[12pt]{article}

\begin{document}

\thispagestyle{empty}

\hfill \parbox{45mm}{{ECT*-99-2} \par March 1999} 

\vspace*{5mm}

\begin{center}
{\LARGE Effect of a scale-dependent cosmological term}

\smallskip

{\LARGE on the motion of small test particles} 

\smallskip

{\LARGE in a Schwarzschild background}
	
\vspace{12mm}

{\large Giovanni Modanese}
\footnote{e-mail: modanese@science.unitn.it}

\medskip        

{\em European Centre for Theoretical Studies in Nuclear Physics
and Related Areas \par
Villa Tambosi, Strada della Tabarelle 286 \par
I-38050 Villazzano (TN) - Italy}

{and}

\medskip

{\em I.N.F.N.\ -- Gruppo Collegato di Trento \par
Dipartimento di Fisica dell'Universit\`a \par
I-38050 Povo (TN) - Italy}
	
\end{center}

\vspace*{5mm}

\begin{abstract}
It was recently suggested that the gravitational action could contain a
scale-dependent cosmological term, depending on the length or momentum 
scale characteristic of the processes under consideration. In this
work we explore a simple possible consequence of this assumption.
We compute the field generated in empty space by a static spherical source 
(the Schwarzschild metric), using the modified action. The resulting 
static potential turns out to contain a tiny non-Newtonian component
which depends on the size of the test particles. The possible relevance 
of this small correction for the analysis of the recent Pioneers data 
[J.D.\ Anderson et al., Phys.\ Rev.\ Lett.\ {\bf 81} (1998) 2858] is 
briefly discussed.

\medskip
\noindent 04.20.-q Classical general relativity.

\noindent 04.60.-m Quantum gravity.

\medskip
\noindent Key words: General Relativity, Quantum Gravity, Experimental 
Gravitation

\bigskip 

\end{abstract}

\section{Introduction}

		On the base of the work by Hamber and Williams in
Euclidean lattice quantum gravity \cite{hw}, it was recently suggested that
the gravitational action could contain a scale-dependent cosmological term
\cite{gm1}. In terms of quantum field theory this means that the effective
value of the cosmological constant would depend on the characteristic
length or momentum scale of the processes under consideration. Obviously
since quantum gravity is not renormalizable, at least in the pure Einstein
formulation, the concept of an effective coupling depending on the scale
is not well defined at the perturbative level. A non perturbative analysis
is needed, like that of Ref.\ \cite{hw}. 

		The issue of a scale-dependent cosmological term is
connected to the well known puzzle of the ``global" cosmological constant
$\Lambda$.  In principle this can be present in the gravitational action
already at the classical level and is expected to receive a large
contribution from the vacuum fluctuations of the quantum fields; still,
its observed value is very close to zero. Several theoretical conjectures
were formulated in order to explain this cancellation \cite{wei}. The
experimental upper limits on a global $\Lambda$ \cite{lim} stem from
observations at cosmological scale and also from astronomical measurements
in the solar system, where a negative $\Lambda$ would correspond to a --
never observed -- Yukawa range for the gravitational potential of the sun
and the planets.

\subsection{The scaling law}

		How does the effective value $\Lambda_{eff}$ of the
cosmological constant (i.e., the average value of the scalar curvature 
evaluated in a certain 4D region) depend on the scale in the quantum
lattice theory? It turns out that $\Lambda_{eff}$ tends to zero when the
size $\xi$ of the region goes to infinity, but for any finite value of
$\xi$ it is given by the power law
	\begin{equation}
\left(|\Lambda_{eff}|G \right)(\xi) \sim (l/\xi)^\gamma
\label{e11}
\end{equation}
	where $l$ is the lattice spacing and $\gamma$ is a critical
exponent. This scale dependence makes sense as long as we regard $l$ as a
small but {\it finite} quantity. In fact in lattice quantum gravity,
unlike in the usual lattice theories, the lattice spacing not only acts as
a regulator, but also represents the minimum physical distance. Thus $l$
cannot be smaller than the Planck length and in the following $l$ will
indeed denote $l_P \sim 10^{-33} \ cm$. (In this work we employ natural
units, such that $\hbar=c=1$. Since in terms of $\hbar$, $c$ and $G$ one
has $l_P \sim \sqrt{G\hbar/c^3}$, the scale dependence of the cosmological
term can be regarded as a quantum effect.) 

		Note that being $l$ extremely small, the ratio $l/\xi$ is
also very small at any reasonable scale. Also note that in order to
isolate $|\Lambda_{eff}|$ in eq.\ (\ref{e11}) we must do some hypothesis
on the behavior of the effective Newton constant $G$. It is generally
believed -- and the numerical simulations confirm this to some extent --
that the scale dependence of $G$ is quite weak. Thus after extracting
$|\Lambda_{eff}|$ from the adimensional product $|\Lambda_{eff}|G$ one
finds that $|\Lambda_{eff}|$ varies essentially as $(l/\xi)^\gamma$.

		The critical exponent $\gamma$ has been numerically
estimated in the simulations, but only for small lattices. Some
consistency arguments (compare Section 2) allow to conclude that $\gamma$
is slightly larger than 2. The simulations also show that the sign of
$\Lambda_{eff}$ is negative: when the size of the 4D region, where the
average of the effective value of $\Lambda$ is computed, tends to
infinity, $\Lambda_{eff}$ tends to zero from below. 

	This feature is familiar in numerical quantum gravity on the Regge
lattice, where a positive value of the effective $\Lambda$ always
corresponds to the unstable, collapsed phase of the system. It is also
connected, in the weak field limit, to a property of the linearized
theory, where a positive cosmological term corresponds to an imaginary
mass for the graviton and thus to unstable solutions.

	Although most of the considerations above refer (like the
numerical simulations) to the Euclidean field theory, the assumption is
made that in the weak field limit this is equivalent to the theory in
Minkowski space.  However, even if we do not trust this equivalence, we
can regard eq.\ (\ref{e11}) as a reasonable {\it ansatz} for the scale
dependence.\footnote{Besides the numerical simulations, there is a further
argument showing that a small negative local cosmological term with the
scaling law (\ref{e11})  is necessary for the stability of the
gravitational vacuum: it stabilizes the Einstein action with respect to a
special set of field configurations, called ``zero-modes" \cite{zer}. This
holds in the Euclidean as well as in the Minkowskian formalism.}

\subsection{Physical implications}

		Suppose we accept the possibility of a scale-dependent
cosmological term, stronger at short distances than at large distances. 
The next step is to understand the physical implications of this formal
assumption. Which physical processes could be affected by such a term? How
can it be taken into account? 

	In an earlier work we studied whether a local cosmological term
could let a graviton decay into collinear gravitons of smaller
frequencies. The decay of a massless particle is a quite exotic process,
whose amplitude obeys certain general rules \cite{dec}; one of these is
that decay is possible if the massless field self-interacts with a
coupling of positive mass dimension (like $\Lambda$ for the gravitons).
However, for any field model comprising a scale-dependent coupling, the
separation of long distance and short distance effects is not trivial. In
the case of gravity this problem is even more serious, because while
speaking of gravitons as of elementary particles we rely on the concept of
Lorentz symmetry; but at the same time we admit that a decay of these
particles could be caused by a non vanishing cosmological term, which
corresponds classically to a curved background. Therefore it was suggested
to consider separately (i)  the phase space for the decay products at
large distance (referred to flat spacetime), and (ii) a local decay
amplitude (where the effective value of $\Lambda$ would appear).  This
program, however, faces too many technical difficulties and
inconsistencies to allow a clear resolution.

		This work aims at investigating the possible effects of a
scale-dependent cosmological term in a simpler, {\it classical} context. 
In Sections 3 and 4 we study the motion of a body with given finite size
in a background gravitational field, under the assumption that this body
also ``feels" a nonzero local cosmological constant, depending on the body
size as in eq.\ (\ref{e11}). We wonder whether the motion is affected, and
to what extent. 

		There are some paradoxes and physical mistakes which lie
in wait for us along this calculation. The first, serious problem is the
violation of the equivalence principle. In the presence of purely
gravitational forces, any dependence of the center of mass motion of a
test body on its size represents such a violation and could be in
principle detected by an E\"otv\"os-like experiment.\footnote{According 
to eq.\ (\ref{e11}) and the following discussion, the scale
dependence of $\Lambda_{eff}$ is a quantum effect, an ``imprint" of
the Planck length physics. One could dispute, at this point, how the
quantization of a theory intrinsically invariant under diffeomorphisms can
lead to results in contrast with the equivalence principle. In fact, as
already discussed by Hamber and Williams, the quantum version of General
Relativity on the Regge lattice reproduces the complete diffeomorphisms
invariance only in the long distance limit.} Decisive points are,
of course, the magnitude of the violation and its exact dependence on the
test body size. This dependence could be so weak to lead to negligible
differences in the motion, except for bodies differing in size by several
magnitude orders.\footnote{The experimental data of Ref.\ \cite{nie} are 
quite inspiring under this respect. Compare our discussion in Section 5.}

		A further crucial problem is a possible violation of the
Lorentz symmetry of free motion. In an empty space, how could the motion
of a test body be affected by the local $\Lambda$-term without spoiling
Lorentz symmetry?  Certainly the body cannot ``slow down" or ``turn" in some
direction. Our calculations show that Lorentz symmetry is actually
preserved and that the $\Lambda$-term only has an effect in the presence
of a non-flat background defining a preferred direction. 

		In order to fix the ideas we chose a Schwarzschild
background metric, i.e., the static spherically symmetric field generated
by a certain mass $M$. We found an expression for the modification of this
metric due to the inclusion of local cosmological term in the
gravitational action, and studied the motion of a test body in the
resulting field. We concluded that the main effect of the cosmological
term is just a small unobservable ``renormalization" of the product $GM$,
independent on the test body size; in addition there is, however, a tiny
non-Newtonian force, larger for smaller bodies and depending on the
velocity $v(r)$ (but unrelated to the familiar post-Newtonian corrections).

		The discussion of this puzzling result is the subject of
Section 5.

\section{The critical exponent in the $\Lambda_{eff}(\xi)$ law}

		Before embarking in the main calculation, let us derive in
this section an estimate of the exponent $\gamma$ in eq.\ (\ref{e11}).
This is obtained just imposing a phenomenological consistency condition.
From eq.\ (\ref{e11}) we can isolate $|\Lambda_{eff}|(\xi)$, obtaining
	\begin{equation}
|\Lambda_{eff}|(\xi) \sim l^{-2} \xi^{-\gamma} l^\gamma .
\label{e21}
\end{equation}
		The range $\rho$ of a Yukawa component in the gravitational
potential is proportional to $|\Lambda_{eff}|^{- 2}$. We thus have 
(up to a numerical factor of order 1 which is already present in eq.\
(\ref{e11}))
	\begin{equation}
\rho \sim l^{1-\gamma/2} \xi^{\gamma/2} .
\label{e22}
\end{equation}
		We see from this equation that if $\gamma$ was exactly
equal to 2, then $\rho=\xi$, i.e.\ the range of the Yukawa component of
the gravitational field observed at the scale $\xi$, would be equal to
$\xi$.  But we know that such a Yukawa component is never observed, thus
we must have $\rho \gg \xi$. For instance, by observing the planetary
motion we can set an upper bound on the average curvature of spacetime at
a scale of the order of the solar system size.

		We therefore assume that $\gamma$ is slightly larger than
2 and define $\alpha=\gamma/2-1$, where $\alpha$ is a small positive
number (of the order of 0.03 or less -- see below). In this way, eq.\
(\ref{e22})  becomes
	\begin{equation}
\rho \sim l^{-\alpha} \xi^{1+\alpha }
\label{e23}
\end{equation}
and the ratio between $\rho$ and $\xi$ is given by
	\begin{equation}
\frac{\rho}{\xi} \sim \left( \frac{\xi}{l} \right)^\alpha .
\label{e24}
\end{equation}
		This ratio does not actually need to be very large,
because the ratio between a possible Yukawa term $V_Y$ in the
gravitational potential and the usual Newton term $V_N$ reduces to an
exponential: 
	\begin{equation}
\frac{V_Y}{V_N} \sim \exp \left(\frac{\xi}{\rho} \right) .
\label{e25}
\end{equation}
	Thus $\rho/\xi$ is the logarithm of the ratio between the upper 
bound on $V_Y$ and $V_N$. Taking, for instance, $V_Y/V_N=10^{-9}$, one
finds $\rho/\xi=\ln(10^9) \simeq 21$. The size $\xi$ of the solar system
is of the order of $\xi=10^{10} \ Km=10^{15} \ cm$, thus the ratio $\xi/l$
in eq.\ (\ref{e24}) is of the order of $10^{48}$. By requiring
$(10^{48})^\alpha < 21$, one finds $\alpha < 0.03$ and $2 < \gamma < 2.03$.

\section{The correction to the Schwarzschild metric}

	In this section we compute to lowest order in the weak field
approximation the field generated by a pointlike static source, 
assuming that the field equations contain a cosmological term localized
at a given point.

\subsection{Einstein field equations and harmonic gauge propagator}

		Let us start by fixing our conventions. The Einstein
equations in the absence of a cosmological term are
	\begin{equation}
R_{\mu \nu}(x) - \frac{1}{2} g_{\mu \nu}(x) R(x) = - 8\pi G T_{\mu \nu}(x). 
\label{e31}
\end{equation}
	The trace of this equation is $R(x) = 8\pi G
T(x)$, where $T$ is the trace of $T_{\mu \nu}$. 

		We focus on the case of a weak field $g_{\mu \nu}(x) 
=\eta_{\mu \nu}+h_{\mu \nu}(x)$; neglecting terms quadratic in $h$, one
obtains the usual linearized version of Einstein equations, which in
harmonic gauge can be written in the form
	\begin{equation}
K_{\mu \nu \rho \sigma}(x) h^{\rho \sigma}(x) = T_{\mu \nu}(x)
\label{e33}
\end{equation}
	where $K$ is a linear differential operator. Given the source 
$T_{\mu \nu}$, the solution of this equation is
	\begin{equation}
h_{\mu \nu}(x) = \int d^4y \, P_{\mu \nu \rho \sigma}(x,y) T^{\rho
\sigma}(y)
\label{e34}
\end{equation}
	where $P$ is the propagator, that is, the inverse operator of $K$, 
given by
	\begin{equation}
P_{\mu \nu \rho \sigma}(x,y) = -\frac{2G}{\pi} 
\frac{Q_{\mu \nu \rho \sigma}}{(x-y)^2-i\varepsilon} 
\label{e35}
\end{equation}
	with $Q$ a constant tensor: 
	\begin{equation}
Q_{\mu \nu \rho \sigma}=\eta_{\mu \rho}\eta_{\nu \sigma}+\eta_{\mu \sigma}
\eta_{\nu \rho}-\eta_{\mu \nu}\eta_{\rho \sigma} .
\end{equation}

		In all these equations it is understood that, in
accordance with the linearized approximation, the indices are lowered and
raised with the metric $\eta_{\mu \nu}$, which is diagonal and has
signature (-1,1,1,1). The square of a 4-vector $x$ is $x^2=-(x^0)^2+{\bf
x}^2$.

\subsection{Field equations with a pointlike source and a local
cosmological term}

		In the presence of a cosmological constant 
$\Lambda$ independent on $x$, eq.\ (\ref{e31}) takes the form 
	\begin{equation}
R_{\mu \nu}(x) - \frac{1}{2} g_{\mu \nu}(x) R(x) + \frac{1}{2} 
\Lambda g_{\mu \nu}(x) = - 8\pi G T_{\mu \nu}(x) .
\label{e32}
\end{equation}
	If $T_{\mu \nu}=0$, this equation admits a solution with
constant scalar curvature $R(x)=2\Lambda$.

	The linearized version of eq.\ (\ref{e32}) is
	\begin{equation}
-8\pi G K_{\mu \nu \rho \sigma}(x) h^{\rho \sigma}(x) + \frac{1}{2} 
\Lambda \left[ \eta_{\mu \nu}+h_{\mu \nu}(x) \right] = -8\pi G T_{\mu
\nu}(x) .
\label{e36}
\end{equation}

		We want now to solve this equation in the case when the
energy-momentum tensor is that of a pointlike mass placed at the origin of
the coordinates, called $T^S_{\mu \nu}(x)$, and the effective cosmological
constant depends on $x$ ($\Lambda \to \Lambda_{eff}(x)$). Let us decompose 
the metric $h_{\mu \nu}(x)$ as follows: 
	\begin{equation}
h_{\mu \nu}(x) = h^S_{\mu \nu}(x) + \Lambda_{eff}(x)h^\Lambda_{\mu \nu}(x)
\label{e37}
\end{equation}
	where the metric $h^S$ satisfies the equation
	\begin{equation}
K_{\mu \nu \rho \sigma}(x) h^{S,\rho \sigma}(x) = T^S_{\mu \nu}(x)
\label{e38}
\end{equation}
	and can be therefore regarded as the linearized version, in
harmonic gauge, of the Schwarzschild metric. (We shall derive $h^S$ in a
moment.)  By introducing the decomposition (\ref{e37}) into eq.\
(\ref{e36}) we obtain
	\begin{eqnarray}
& & -8\pi G K_{\mu \nu \rho \sigma}(x) \left[h^{S,\rho \sigma}(x) + 
\Lambda_{eff}(x) h^{\Lambda,\rho \sigma}(x) \right] + \nonumber \\
& & + \frac{1}{2} \Lambda_{eff}(x) \left[ \eta_{\mu \nu} + h^S_{\mu
\nu}(x) + 
\Lambda_{eff}(x) h^\Lambda_{\mu \nu}(x) \right] = -8\pi G T^S_{\mu \nu}(x).
\label{e39}
\end{eqnarray}

		Now let us write the function $\Lambda_{eff}(x)$ in the
form $\Lambda_{eff}(x)=\Lambda_{eff} f(x)$, where $f(x)$ is an
adimensional function vanishing outside a region of width $d$ (size of the 
test particle) centered at ${\bf X}$ (center of mass coordinate of the test 
particle), and such that $f({\bf X})=1$. In eq.\ (\ref{e39}) we can single
out terms of zeroth order in $\Lambda_{eff}$, which cancel out due to eq.\
(\ref{e38}), one term of order $\Lambda_{eff}^2$, which can be neglected,
and terms of the first order in $\Lambda_{eff}$, which lead to the
following equation for the metric $h^\Lambda$: 
	\begin{equation}
-8\pi G K_{\mu \nu \rho \sigma}(x) f(x) h^{\Lambda,\rho \sigma}(x) + 
\frac{1}{2} f(x) \left[\eta_{\mu \nu} + h^S_{\mu \nu}(x) \right] = 0.
\label{e310}
\end{equation}

		If we know the background metric $h^S(x)$, we can find the
correction $h^\Lambda(x)$ due to the cosmological term in the neighbourhood
of ${\bf X}$ by applying the operator $P$, inverse of $K$: 
	\begin{equation}
h^\Lambda_{\mu \nu}(x) = \frac{1}{8\pi G f(x)} \int d^4y \, 
P_{\mu \nu \rho \sigma}(x,y) \frac{1}{2} f(y) \left[ \eta^{\rho \sigma} + 
h^{S,\rho \sigma}(y) \right].
\label{e311}
\end{equation}

\subsection{Computation of $h^S$}

		In order to find $h^S(x)$, we replace $T^S(x)$ by its
explicit expression
	\begin{equation}
T^S(x) = M \delta^3({\bf x}) \delta_{\mu 0} \delta_{\nu 0}
\label{e312}
\end{equation}
	and integrate eq.\ (\ref{e33}):
	\begin{eqnarray}
h^S_{\mu \nu}(x) & = & \int d^4y \, P_{\mu \nu \rho \sigma}(x,y) M 
\delta^3({\bf y}) \delta^{\rho 0} \delta^{\sigma 0} = \nonumber \\
& = & M \int dy^0 P_{\mu \nu 00}(x;y_0,{\bf y}=0) = \nonumber \\
& = & M \frac{-2G}{\pi} \left( 2\eta_{\mu 0}\eta_{\nu 0}
-\eta_{\mu \nu}\eta_{00} \right) \int dy^0 \frac{1}{{\bf x}^2-(x^0-y^0)^2
-i\varepsilon} = \nonumber \\
& = & \frac{2GM}{|{\bf x}|} \left( 2\eta_{\mu 0}\eta_{\nu 0}-
\eta_{\mu \nu}\eta_{00} \right).
\label{e313}
\end{eqnarray}

		Then, in conclusion, $h^S$ is given by
	\begin{equation}
h^S_{00}({\bf x}) = \frac{2GM}{|{\bf x}|} (\eta_{00})^2 = 
\frac{2GM}{|{\bf x}|}
\label{e314}
\end{equation}
	which is correct, since in general in a static field one has
$h_{00}=-2V_{Newtonian}$; the $ii$ components are the same:
	\begin{equation}
h^S_{ii}({\bf x}) = \frac{2GM}{|{\bf x}|} (-\eta_{ii} \eta_{00}) =
\frac{2GM}{|{\bf x}|}.
\label{e315}
\end{equation}

\subsection{Computation of $h^\Lambda$}

		Now we can compute $h^\Lambda$. In particular, we are
interested into the 00 component, which gives the correction to the static
potential felt by the test particle. We evaluate it at the center of mass
coordinate $X=(X_0,{\bf X})$ and find
	\begin{eqnarray}
h^\Lambda_{00}(X) &= & \frac{1}{8\pi G f(X)} \int d^4y \, P_{00 \rho
\sigma}(X,y)
\frac{1}{2} f(y) \left[\eta^{\rho 
\sigma} + h^{S,\rho \sigma}(y) \right] \nonumber \\
&= & -\frac{1}{8\pi^2} \int d^4y \, \frac{2\eta_{0\rho} \eta_{0\sigma}
-\eta_{00}
\eta_{\rho \sigma}}{(X-y)^2-i\varepsilon} 
f(y) \left[\eta^{\rho \sigma} + h^{S,\rho \sigma}(y) \right] .
\label{e316}
\end{eqnarray}
		Before looking at the integral, let us focus on the
algebraic part of this expression. The term $\eta^{\rho \sigma}$ in the
bracket $\left[\eta^{\rho \sigma} + h^{S,\rho \sigma}(y)  \right]$, after
multiplication by $\left( 2\eta_{0\rho} \eta_{0\sigma} -\eta_{00}
\eta_{\rho \sigma}\right)$, gives
	\begin{equation}
\left( 2\eta_{0\rho} \eta_{0\sigma} -\eta_{00} \eta_{\rho \sigma}\right)
\eta^{\rho \sigma} = 2\eta_{00}-4\eta_{00} = -2+4 
= 2.
\end{equation}
	In a similar way, for the term $h^{S,\rho \sigma}(y)$ we obtain
	\begin{eqnarray}
& & \left( 2\eta_{0\rho} \eta_{0\sigma} -\eta_{00} \eta_{\rho \sigma}\right)
h^{S,\rho \sigma}(y) = \nonumber \\
& & \qquad \qquad = 2h^S_{00}(y) - (\eta_{00})^2 h^S_{00}(y) - \eta_{00} \sum_i
h^S_{ii}(y) = \nonumber \\
& & \qquad \qquad = \frac{2GM}{|{\bf y}|} 
(2-1+3) = \frac{8GM}{|{\bf y}|}.
\end{eqnarray}
		Now let us write the function $f$ in the following form 
	\begin{equation}
f(y) = 1 \ if \  |{\bf y}-{\bf X}| < d; \qquad f(y)=0 \ elsewhere
\label{e317}
\end{equation}
	or
	\begin{equation}
f(y) = \theta(d-|{\bf y}-{\bf X}|).
\label{e318}
\end{equation}
	Setting for simplicity $X_0=0$, since everything is static, we
obtain the correction
	\begin{equation}
h^\Lambda_{00}({\bf X}) = -\frac{1}{8\pi^2} \int d^4y 
\frac{\theta(d-|{\bf y}-{\bf X}|)}{({\bf y}-{\bf X})^2-(y_0)^2-
i\varepsilon} \left(2+\frac{8GM}{|{\bf y}|} \right) .
\label{e319}
\end{equation}
		Introducing the new integration variable ${\bf z}=
{\bf X}-{\bf y}$, we obtain
	\begin{equation}
h^\Lambda_{00}({\bf X}) = -\frac{1}{8\pi^2} \int dy_0 \int d{\bf z}
\frac{\theta(d-|{\bf z}|)}{{\bf z}^2-(y_0)^2-i\varepsilon} 
\left(2+\frac{8GM}{|{\bf z}-{\bf X}|} \right) .
\label{e320}
\end{equation}
	By integrating the first term in the bracket $(2+8GM/|{\bf z}-{\bf
X}|)$ one clearly obtains a constant independent from ${\bf X}$; this
constant is irrelevant in the present case, since it does not give any
contribution to the force. (This was expected: in flat space there cannot
be any force on the test particle due to the local cosmological term,
otherwise Lorentz invariance would be spoiled.) 
	From the second term in the bracket $(2+8GM/|{\bf z}-{\bf X}|)$ we
obtain
	\begin{equation}
h^\Lambda_{00}({\bf X}) = -\frac{GM}{\pi^2} \int dy_0 \int d{\bf z} 
\frac{\theta(d-|{\bf z}|)}{{\bf z}^2-(y_0)^2-i\varepsilon} 
\frac{1}{|{\bf z}-{\bf X}|} .
\label{e321}
\end{equation}
		Since
	\begin{equation}
\int dy_0 \frac{1}{{\bf z}^2-(y_0)^2-i\varepsilon} = - \frac{\pi}{|{\bf z}|}
\end{equation}
	(compare the calculation of $h^S$), we finally find
	\begin{equation}
h^\Lambda_{00}({\bf X}) = \frac{GM}{\pi} \int d{\bf z} 
\frac{\theta(d-|{\bf z}|)}{|{\bf z}||{\bf z}-{\bf X}|} .
\label{e323}
\end{equation}
	This integral is well known: apart from the factor $-1/\pi$, it is
the Newtonian potential generated at the point ${\bf X}$ by a mass
distribution with spherical symmetry centered at the origin, having
density $\rho(|{\bf z}|)=\theta(d-|{\bf z}|)/|{\bf z}|$.  The integral is
convergent, positive, and proportional to $1/|{\bf X}|$; the
proportionality constant is a pure number of order 1 multiplied by $d^2$.

	Therefore
	\begin{equation}
h^\Lambda_{00}({\bf X}) \sim  \frac{GMd^2}{|{\bf X}|} .
\label{e324}
\end{equation}
		We recall that the static potential is $V=-h_{00}/2$. 
The correction to the Schwarzschild metric is
$\Lambda_{eff}(x)h^\Lambda_{\mu \nu}(x)$, which computed at ${\bf X}$
gives a correction to the potential of the form
	\begin{equation}
V^\Lambda({\bf X}) \sim - \frac{GM\Lambda_{eff} d^2}{|{\bf X}|}
\label{e325}
\end{equation}
	where $\Lambda_{eff}<0$ and ``$\sim$" means that an adimensional
constant of order 1 is omitted.

\section{The factor $|\Lambda_{eff}| d^2$}

	Eq.\ (\ref{e325}) gives the first order correction to the 00
component of the Schwarzschild metric, due to a local cosmological term
felt by the test particles. This correction is proportional to the product
$|\Lambda_{eff}| d^2$, where $d$ is the size of the test particles. With
the notation of Section 2, and supposing for the moment that the length
scale $\xi$ which fixes the value of $\Lambda_{eff}$ corresponds simply to
$d$, the product $|\Lambda_{eff}| d^2$ can also be written as $(d/\rho)^2$
and is thus of order $(l/d)^\alpha$ (eq.\ (\ref{e24})).

	The ratio $(l/d)$ is very small, but since the exponent $\alpha$
is small, too, $(l/d)^\alpha$ turns out to be a ``reasonable" number also
at a macroscopic scale; for instance, if $d=1\ cm$, then $(l/d)^\alpha
\sim (10^{-33})^{0.03} \sim 10^{-1}$. Therefore the correction
(\ref{e325}) is comparable in magnitude to the Newtonian component $-GM/r$
(we set $|{\bf X}|\equiv r$ in the following). In fact, the two
contributions to $h_{00}$ are indistinguishable, since they have the same
$r$-dependence, or in other words the cosmological correction just seems
to lead to a renormalization of the factor $GM$. 

\subsection{$\xi$ depends on the test particles' velocity}

	The simple identification $\xi=d$ is not entirely correct,
however. In the following, we show that $\xi$ has a weak dependence on the
velocity of the test particle, and thus the $r$-dependence of the
cosmological correction is not exactly $1/r$. We first recall that the
value of $\Lambda_{eff}$ is the result of a spatial average. Now, suppose
to follow the motion of the test body for a short time interval $\Delta
t$; the scale $\xi$ corresponds to the cubic root of the volume swept by
the body in this interval, namely
	\begin{equation}
\xi \sim \left( d^3 + d^2v\Delta t \right)^{1/3} .
\label{e41}
\end{equation}
	We see that $\xi$ depends also on the product $v\Delta t$. For
very large bodies (for instance, planets) one can consistently let $\Delta
t$ approach zero, so that the $d^3$ term in the bracket dominates, and
$\xi=d$ with good approximation. For small test bodies, however, the term
$d^2v\Delta t$ will dominate, unless $\Delta t$ is so small to correspond
to an impracticable observation (for instance, if $d=0.1 \ m$ and $v=10^4
\ m/s$, $\Delta t$ should be less than $10^{-5} \ s$). In other words,
even though we are looking for the contribution of the local cosmological
term to the potential $V(r)$ at a given point, for small bodies 
$\Lambda_{eff}$ depends on the volume swept in the motion during a short 
time interval and thus also on the velocity $v$. 

\subsection{Dependence of $V^\Lambda$ on $v(r)$}

	Let us go back to the expression (\ref{e21}) for $\Lambda_{eff}$,
namely $|\Lambda_{eff}| \sim \xi^{- 2}$ (in this computation we can
neglect $\alpha$ as compared to 2, thus $\gamma \simeq 2$). We set $\xi
\sim \left( d^3 + d^2v\Delta t \right)^{1/3}$ like in eq.\ (\ref{e41})
and focus on the motion of a small test body, starting from a certain 
distance $r_0$. Disregarding the term $d^3$ in comparison to $d^2v\Delta t$
we find that during this motion the adimensional ratio
$|\Lambda_{eff}(r)|/|\Lambda_{eff}(r_0)|$ is determined by the ratio of
the velocities at the distances $r$ and $r_0$: 
	\begin{equation}
\frac{|\Lambda_{eff}(r)|}{|\Lambda_{eff}(r_0)|} = \left( \frac{v(r)}{v(r_0)}
\right)^{-2/3} .
\label{e42}
\end{equation}
		Note that $\Delta t$ has dropped from this expression, as
one should expect. Inserting it into eq.\ (\ref{e325}) we obtain: 
	\begin{equation}
V^\Lambda(r) \sim \frac{GM}{r} \left[ d^2 |\Lambda_{eff}(r_0)| 
\right] \left(
\frac{v(r)}{v(r_0)} \right)^{-2/3} .
\label{e43}
\end{equation}

	According to the remarks at the beginning of this section, the
constant factor $d^2 |\Lambda_{eff}(r_0)|$ is of order $\sim 1$. The
$r$-dependence of the factor $(v(r)/v(r_0))^{-2/3}$ causes a deviation
from the pure $1/r$ behavior. This dependence can be obtained taking into
account the total energy conservation. In the present approximation it
suffices to consider the usual non relativistic conservation law: 
	\begin{equation}
-\frac{mMG}{r} + \frac{1}{2} mv^2(r) = -\frac{mMG}{r_0} + \frac{1}{2} mv^2(r_0)
\label{e44}
\end{equation}
	or, denoting $v^2(r_0)$ by $v^2_0$
	\begin{equation}
\frac{v^2(r)}{v^2_0} = 1 + \frac{2MG}{v^2_0} \left( \frac{r_0-r}{rr_0} 
\right)= 1 +
\frac{2g_0}{v^2_0} \frac{r_0}{r} (r_0-r)
\label{e45}
\end{equation}
	where $g_0$ is the gravitational acceleration at the distance
$r_0$. The constant $g_0/v^2_0$ depends on the specific situation. For
instance, the field of the sun at distance $r_0 = 1 \ AU$ is $g_0
\simeq 10^{-3} \ m/s^2$, and taking as velocity of the test object $v_0
\sim 10^4 \ m/s$ one obtains $g_0/v^2_0 \simeq 10^{-11}\ m^{-1}$.

	For small displacements from $r_0$, the general $r$-dependence in
eq.\ (\ref{e45}) can be expanded in powers of $(r-r_0)$ as follows
($b \equiv g_0/v_0^2$)
	\begin{equation}
\left( \frac{v(r)}{v(r_0)} \right)^{-2/3} \simeq 1 + \frac{1}{3} b (r-r_0) +
\frac{1}{9} b \left( 2b-\frac{3}{r_0} \right)  (r-
r_0)^2 + ...
\label{e46}
\end{equation}
	After inserting this expression into eq.\ (\ref{e43}), one finds: 
(i) a potential $1/r$, which corresponds as noticed above to an invisible
``renormalization" of the product $MG$; (ii) an irrelevant constant
potential; (iii) a potential linear in $r$, namely
	\begin{equation}
V^\Lambda_{Lin}(r) \sim GM \left[ d^2 |\Lambda_{eff}(r_0)| \right] b 
\left( 2b-\frac{3}{r_0} \right) r .
\label{e47}
\end{equation}
	Carrying on the expansion (\ref{e46}) to higher orders, one gets
potentials growing like $r^2$, $r^3$... This holds, however, only for $r$
very close to $r_0$. In general, the expression (\ref{e45}) should be
used.

\section{Discussion}

	In the previous sections we investigated some consequences of the
insertion into Einstein field equations of a local, scale-dependent
cosmological term. By this we mean an effective cosmological constant
$\Lambda_{eff}$ (essentially due to the quantum fluctuations of the
gravitational field) whose strength depends on the size of the involved
spatial region. 

	For instance, if we consider the motion of a test particle in a
given background field, the effective cosmological term felt by the
particle depends on the volume swept by the particle in a short time
interval during its motion. We computed the contribution of this term in
the case of a Schwarzschild background, and found that the potential $V=-
h_{00}/2$ acting on a small test particle, usually of the Newtonian form
$V=-GM/r$, is corrected by a small term depending on $r$ as
$v^{-2/3}(r)/r$, being $v(r)$ the particle velocity. 

	The notion of a local cosmological term which ``escorts" a
particle in its motion is consistent with the definition of
$\Lambda_{eff}$ in lattice quantum gravity as the average curvature
measured in a region of given size. However, the extrapolation of this
notion from lattice theory and its implementation in the field equations
could be properly justified only within a yet-to-come full quantum field
theory of gravity. The situation reminds certain issues from the pre-QFT
years early in this century, when quantum concepts were inserted by hand
into classical field theory. Our present approach allows, in some sense,
to take into account the effect of the quantum gravitational fluctuations
on a classical curved background.

	The conclusions of our computation, namely the existence of a tiny
non-Newtonian component of the field depending on the size of the test
particles, can be related to recent experimental results \cite{nie}
which called into play such a small ``unmodeled" gravitational acceleration to
explain the fine-detail motion of some space probes (while the planetary
motion would not be affected). 

	However our phenomenological approach does not allow to fix the
numerical proportionality constants appearing in eq.s (\ref{e43}),
(\ref{e47}). Furthermore, the power expansion in eq.\ (\ref{e47}) is not
suitable for direct comparison with the experimental data. These originate
from a complex fitting procedure, applied to the acceleration measured
over long paths and based on conventional subtraction of the known parts
of the field (with $GM$ obviously constant). If the results of Ref.\
\cite{nie} will be confirmed, it may be worth to consider the inclusion of
the potential (\ref{e43}) in these fits.

\bigskip

This work was partially supported by the A.S.P. -- Associazione per lo 
sviluppo scientifico e tecnologico del Piemonte, Torino, Italy.

\end{document}